\documentclass[twocolumn,showpacs,prb,amsfonts,amsmath,amssymb]{revtex4}
\usepackage{mathrsfs}
\usepackage{graphicx}
\usepackage{array}
\usepackage{bm}
\usepackage[dvipdfm]{hyperref}
\usepackage{mathptmx}

\begin{document}
\title{Spin states and persistent currents in a mesoscopic ring with an embedded magnetic impurity}
\author{J. S. Sheng}
\author{Kai Chang}
\email{[Electronic address: ]kchang@red.semi.ac.cn}
\affiliation{SKLSM, Institute of Semiconductors, Chinese Academy of Sciences, P. O. Box
912, Beijing 100083, China}

\begin{abstract}
Spin states and persistent currents are investigated theoretically
in a mesoscopic ring with an embedded magnetic ion under a uniform
magnetic field including the spin-orbit interactions. The magnetic
impurity acts as a spin-dependent $\delta$-potential for electrons
and results in gaps in the energy spectrum, consequently suppresses
the oscillation of the persistent currents. The competition between
the Zeeman splittings and the $s$-$d$ exchange interaction leads to
a transition of the electron ground state in the ring. The interplay
between the periodic potential induced by the Rashba and Dresselhaus
spin-orbit interactions and the $\delta$-potential induced by the
magnetic impurity leads to significant variation in the energy
spectrum, charge density distribution, and persistent currents of
electrons in the ring.
\end{abstract}

\pacs{73.23.Ra,71.70.Ej,71.70.Gm,75.50.Pp}
\maketitle

\section{INTRODUCTION}\label{sec:introduction}

Recently, spin dependent optical and transport properties of
semiconductors have received renewed interest due to the potential
applications of spintronic
devices.~\cite{Wolf-2001-1488,Zutic-2004-323} One of the essential
requirements in a spintronic device is to generate spin-polarized
current. In a diluted magnetic semiconductor (DMS), the $s$-$d$
exchange interaction~\cite{Furdyna-1988-R29} between the electrons
and the magnetic impurities makes it possible to tailor electron
spin splitting and thereby generate spin-polarized
currents~\cite{Fiederling-1999-787,Ohno-1999-790,Jaroszynski-1995-3170,Sawicki-1986-508}.

The spin states of the electrons in a semiconductor can also be
manipulated by an external electric field via the spin-orbit
interaction (SOI). There are two types of SOI's in semiconductors.
One is the Rashba spin-orbit interaction (RSOI) induced by structure
inversion asymmetry,~\cite{Rashba-1960-1109,Bychkov-1984-6039} and
the other is the Dresselhaus spin-orbit interaction (DSOI) induced
by bulk inversion asymmetry~\cite{Dresselhaus-1955-580}. The
strength of the RSOI can be tuned by external gate voltages or
asymmetric doping, while the strength of the DSOI is inversely
proportional to the thickness of the quantum well and thus becomes
comparable with that of RSOI in narrow quantum
wells.~\cite{Lommer-1988-728} SOI makes it possible to generate a
spin current (SC) electrically without the use of ferromagnetic
material or a magnetic
field.~\cite{Murakami-2003-1348,Sinova-2004-126603} The impact of
in-plane magnetic field on the spin hall conductivity has been
investigated in the two-dimensional electron gas (2DEG) in the
presence of both the RSOI and DSOI.~\cite{Chang-2005-085315} The
coexisting of the $s$-$d$ exchange interaction and spin-orbit
interactions in 2DEG could result in a significant change of the
spin polarization of the charge current
(CC).~\cite{Yang-2006-132112}

Recent progress in fabrication techniques makes it possible to dope
a few or even one magnetic impurity in a semiconductor
nanostructure.~\cite{Besombes-2004-207403} Aharonov-Bhom
oscillations and spin-polarized transport properties in a mesoscopic
open ring with an embedded magnetic impurity have been
investigated~theoretically~\cite{Joshi-2001-075320} using quantum
wave-guide theory~\cite{Xia-1992-3593} without spin-orbit
interactions. Persistent charge and spin currents in closed
mesoscopic rings in the presence of a nonmagnetic impurity were
studied including the RSOI,~\cite{Splettstoesser-2003-165341} and a
rounding effect of the nonmagnetic impurity on the energy spectrum
and the flux oscillation of the persistent CC was found. However, it
is interesting to investigate the effect of a magnetic impurity on
persistent currents, especially spin current, in a mesoscopic ring.
The interplay between the RSOI and DSOI can induce an effective
azimuthal periodic potential in the ring, consequently breaks the
cylindrical symmetry of the ring,~\cite{Sheng-2006-235315} and this
feature makes the spin states and the persistent currents depend
sensitively on the position of the magnetic impurity.

In this work, we study spin states and persistent currents of a 1D
mesoscopic ring with an embedded magnetic impurity in the presence
of both the RSOI and DSOI. The interplay between the Zeeman
splittings and the $s$-$d$ exchange interaction leads to a
transition of the electron ground state. The energy spectrum and the
persistent currents depend sensitively on the position of the
magnetic impurity including both the RSOI and DSOI, since the
interplay between the RSOI and DSOI breaks the cylinderical
symmetry. It is interesting to notice that the symmetry of the
persistent SC in the parameter space ($\alpha$-$\beta$) is robust
against the magnetic impurity. The paper is organized as follows.
The theoretical model is presented in Sec.~\ref{sec:theory}. The
numerical results and discussion are given in
Sec.~\ref{sec:results}. Finally, we give a brief conclusion in
Sec.~\ref{sec:summary}.

\section{THEORETICAL MODEL}\label{sec:theory}

\begin{figure}[ptbh]
\includegraphics[width=\columnwidth]{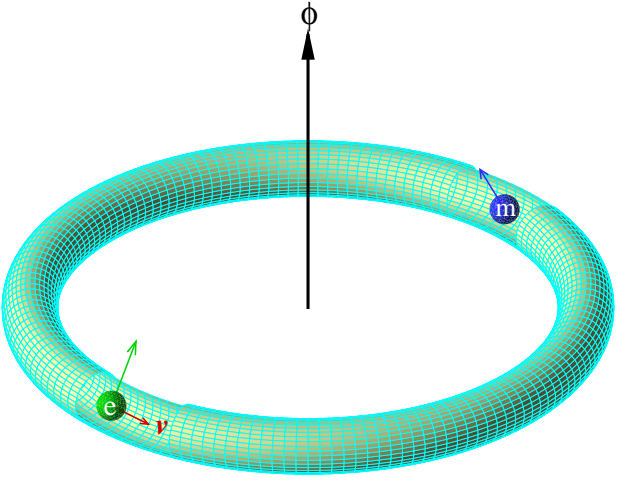}\caption{(Color
online) Schematic diagram for a mesoscopic ring with an embedded
magnetic impurity.} \label{fig:schematic}
\end{figure}

In the presence of both the RSOI and DSOI, the dimensionless
Hamiltonian of a mesoscopic ring with an embedded magnetic impurity
(see Fig.~\ref{fig:schematic}) under a uniform perpendicular
magnetic field reads~\cite{Sheng-2006-235315}
\begin{align}
H &  =\left[  -i\frac{\partial}{\partial
\varphi}+\phi+\frac{\alpha}{2}
\sigma_{r}-\frac{\beta}{2}\sigma_{\varphi}(-\varphi)\right]
^{2}-\frac
{\alpha^{2}+\beta^{2}}{4}\nonumber \\
&  +\frac{\alpha \beta}{2}\sin2\varphi+g_{e}\phi
\sigma_{z}^{e}+g_{m}\phi \sigma_{z}^{m}-2\pi J\hat{\bm{s}}^{e}\cdot
\hat{\bm{s}}^{m}\delta (\varphi-\theta),\label{eqn:hami}
\end{align}
where $\phi$ is the magnetic flux in the units of $\phi_{0}=h/e$,
and $\alpha$ and $\beta$ specify the strengths of the RSOI and DSOI,
respectively. $\sigma _{r}=\cos \varphi \sigma_{x}^{e}+\sin \varphi
\sigma_{y}^{e}$ and $\sigma _{\varphi}=\cos \varphi
\sigma_{y}^{e}-\sin \varphi \sigma_{x}^{e}$, $g_{e}$ ($g_{m}$) is
the $g$ factor of the electron (magnetic impurity), and $J $ is the
strength of the $s$-$d$ exchange interaction between the conduction
band electron $\hat{\bm{s}}^{e}$ and the magnetic impurity
$\hat{\bm{s}}^{m}$. As discussed in our previous work, the interplay
between the RSOI and DSOI induces a $\sin2\varphi$ potential [the
third term in Eq.~(\ref{eqn:hami})] and breaks the cylindrical
symmetry of the mesoscopic ring.~\cite{Sheng-2006-235315}

The charge density operator and the charge current density operator are
\begin{equation}
\begin{split}
\hat{\rho}(\varphi^{\prime}) &  =-e\delta(\varphi^{\prime}-\varphi)\\
\hat{\bm{j}}_{c}(\varphi^{\prime}) &  =\frac{1}{2}[\hat{\rho}(\varphi^{\prime
})\hat{\bm{v}}+\hat{\bm{v}}\hat{\rho}(\varphi^{\prime}))],
\end{split}
\label{eqn:coperators}
\end{equation}
where $\varphi^{\prime}$ refers to the field coordinates and
$\varphi$ the coordinates of the electron. The velocity operator
associated with the Hamiltonian in Eq.~(\ref{eqn:hami}) is
\begin{equation}
\hat{\bm{v}}=\mathbf{e}_{\varphi}\left[  -i\frac{\partial}{\partial
\varphi
}+\phi+\frac{\alpha}{2}\sigma_{r}-\frac{\beta}{2}\sigma_{\varphi}
(-\varphi)\right]  .\label{eqn:velocity}
\end{equation}
We can also introduce the spin density and spin current density operators as
\begin{equation}
\begin{split}
\hat{\bm{S}}(\varphi^{\prime}) &  =\hat{\bm{s}}^{e}\delta(\varphi^{\prime
}-\varphi)\\
\hat{\bm{j}}_{s}(\varphi^{\prime}) &  =\frac{1}{2}[\hat{\bm{S}}(\varphi
^{\prime})\hat{\bm{v}}+\hat{\bm{v}}\hat{\bm{S}}(\varphi^{\prime})],
\end{split}
\label{eqn:soperators}
\end{equation}
where $\hat{\bm{s}}^{e}$ is the vector of the electron spin operator. The
charge current density and spin current density can be obtained by calculating
the expectation values of the corresponding operators:
\begin{equation}
\begin{split}
\bm{j}_{c}(\varphi^{\prime}) &  =\langle \Psi|\hat{\bm{j}}_{c}|\Psi
\rangle=-e\operatorname{Re}\{
\Psi^{\dag}(\varphi^{\prime})\hat{\bm{v}}
^{\prime}\Psi(\varphi^{\prime})\} \\
\bm{j}_{s}(\varphi^{\prime}) &  =\langle \Psi|\hat{\bm{j}}_{s}|\Psi
\rangle=\operatorname{Re}\{ \Psi^{\dag}(\varphi^{\prime})\hat{\bm{v}}^{\prime
}\hat{\bm{s}}^{e}\Psi(\varphi^{\prime})\},
\end{split}
\label{eqn:cdexp}
\end{equation}
where $\Psi(\varphi)$ is the wavefunction of an electron in the
ring. For convenience, we note $\varphi^{\prime} and
\hat{\bm{v}}^{\prime}$ as $\varphi and \hat{\bm{v}}$ hereafter.

The azimuthal (spin or charge) current can be defined
as~\cite{Wendler-1994-4642}
\begin{equation}
I=\frac{1}{2\pi}\int_{0}^{2\pi} \! \! \! \mathrm{d} \varphi
j(\varphi ).\label{eqn:current}
\end{equation}

At low temperature, $N$ electrons will occupy the lowest $N$ levels
of the energy spectrum. The total (charge or spin) current is the
summation over all occupied
levels.~\cite{Splettstoesser-2003-165341}

\section{NUMERICAL RESULTS AND DISCUSSION}\label{sec:results}

\subsection{The effects of the magnetic impurity}

In order to clearly investigate the effects of the magnetic impurity
on the spin states and persistent currents in the 1D ring, we first
neglect the Zeeman splittings and spin-orbit interactions. The
Hamiltonian of the system becomes

\begin{equation}
H=\left( -i\frac{\partial}{\partial \varphi}+\phi \right) ^{2}-2\pi
J \hat{\bm{s}}^{e}\cdot \hat{\bm{s}}^{m}\delta(\varphi-\theta
).\label{eqn:hamilnosoi}
\end{equation}

\begin{figure}[ptbh]
\includegraphics[width=\columnwidth]{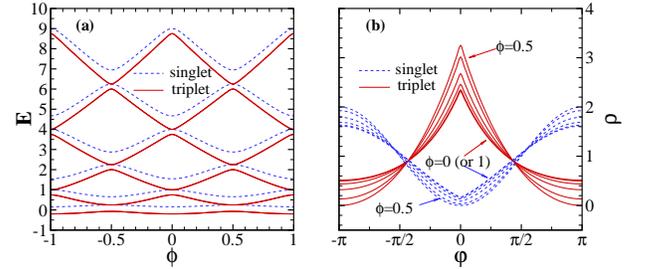}\caption{(Color
online) (a) Energy spectrum of 1D mesoscopic ring with one magnetic
impurity at different magnetic fluxes; (b) Probability density
distribution for the lowest triplet and singlet states of 1D ring
with one magnetic impurity for different magnetic flux. $J=0.5$ and
$\theta=0$.} \label{fig:eigen_pd_nosoi}
\end{figure}

When a spin-1/2 magnetic impurity appears in the mesoscopic ring,
the total angular momentum of the eigenstates is equal to $1$
(triplet) or $0$ (singlet) due to the coupling between the electron
spin $\hat{\bm{s}}^{e}$ and the impurity spin $\hat{\bm{s}}^{m}$. In
the coupling representation, the Hamiltonian can be written as
\begin{equation}
H_{c}=
\begin{bmatrix}
H_{S} & 0\\ 0 & H_{T}I_{3}
\end{bmatrix},\label{eqn:hamic}
\end{equation}
where $H_{S}=\left(  -i\frac{\partial}{\partial \varphi}+\phi
\right) ^{2}+\frac{3\pi J}{2}\delta(\varphi-\theta)$ is the
Hamiltonian for the singlet states ($j=0$), $H_{T}=\left(
-i\frac{\partial}{\partial \varphi} +\phi \right)  ^{2}-\frac{\pi
J}{2}\delta(\varphi-\theta)$ is the Hamiltonian for the triplet
states ($j=1$), and $I_{3}$ represents the $3\times3$ identity
matrix. We can find that the magnetic impurity behaves like a
barrier (well) for the singlet (triplet) state when $J>0$.

Fig.~\ref{fig:eigen_pd_nosoi}(a) shows the energy spectrum of the
mesoscopic ring with an embedded magnetic impurity. The energy
splittings between the triplet and singlet states are proportional
to the strength of the exchange interaction $J>0$. The neighboring
singlet and triplet levels can degenerate at special magnetic fluxes
(integer or half-integer $\phi$). When $\phi$ is an integer, the
energies of the fourfold degenerate levels are $1,4,9,\cdots
,n^{2},\cdots$, which can be obtained from $\sin \pi \kappa=0$ in
both Eq.~(\ref{eqn:sinteger}) and Eq.~(\ref{eqn:trinteger1}) in the
Appendix; When $\phi$ is a half-integer, the energies of the
fourfold degenerate levels are
$1/4,9/4,25/4,\cdots,(n+1/2)^{2},\cdots$, which can be obtained from
$\cos \pi \kappa=0$ in both Eq.~(\ref{eqn:shalfinterger}) and
Eq.~(\ref{eqn:thalfinterger1}) in the Appendix.

Note that the orbital wavefunction and spin wavefunction can be separated in
the ring without SOI. The eigenstates of the system can be written as
$\Psi=X_{jm}(\varphi)|\frac{1}{2}\frac{1}{2}jm\rangle$ where $j=0,1$ and
$m=-j,\ldots,j$. The orbital wavefunctions $X_{jm}$ are determined by
\begin{subequations}
\begin{align}
\left(  -i\frac{\partial}{\partial \varphi}+\phi \right)
^{2}X_{00}+\frac{3}
{2}\pi J\delta(\varphi-\theta)X_{00} &  =EX_{00},\label{eqn:singlet}\\
\left(  -i\frac{\partial}{\partial \varphi}+\phi \right)
^{2}X_{1m}-\frac{1} {2}\pi J\delta(\varphi-\theta)X_{1m} &
=EX_{1m}.\label{eqn:triplet}
\end{align} \label{eqn:singletandtriplet}
\end{subequations}
This means that the spin-1/2 magnetic impurity acts as a
$\delta$-barrier (well) and well (barrier) on the singlet and
triplet states for $J>0$ ($J<0$), respectively. This feature
consequently leads to different localization of the singlet and
triplet states [see Fig.~\ref{fig:eigen_pd_nosoi}(b)]. We assume
$J>0$ in this work without loss of generality. In this case, the
ground state of the system is a triplet state.

\begin{figure}[ptbh]
\includegraphics[width=\columnwidth]{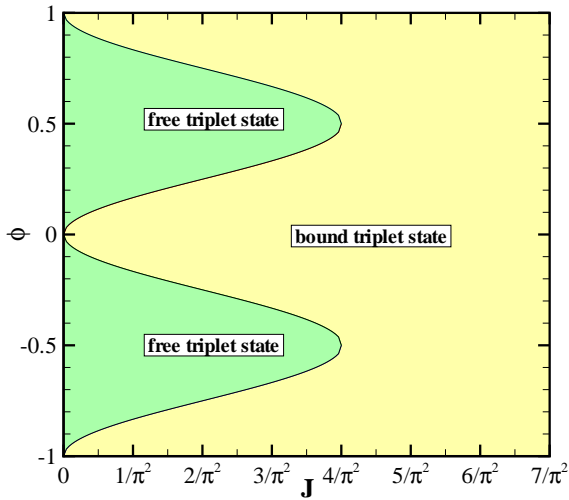}\caption{(Color
online) The phase diagram of the ground triplet state in the ring at
different magnetic flux $\phi$ and $s$-$d$ exchange interaction
strength $J$.} \label{fig:free_or_bound}
\end{figure}

We now focus on the lowest triplet state. The transition between
bound state and free state can be clearly seen in
Fig.~\ref{fig:free_or_bound} as a function of the magnetic flux
$\phi$\ and the strength of the $s$-$d$ exchange interaction $J$.
The lowest triplet state is a bound state at an integer $\phi$
because Eq.~(\ref{eqn:trinteger2}) has one solution for any given
positive $J$. Whether the lowest triplet state at a half-integer
$\phi$ is a bound state ($E<0$) or free state ($E>0$) depends on the
strength of the $s$-$d$ exchange interaction $J$. When
$J>4/\pi^{2}$, Eq.~(\ref{eqn:thalfinterger2}) has a nontrivial
solution, and therefore the lowest triplet state at a half-integer
$\phi$ is a bound state. When $J<4/\pi^{2}$,
Eq.~(\ref{eqn:thalfinterger2}) only has a trivial solution, and thus
the lowest triplet state at a half -integer $\phi$ is a free state.
This means that when the strength of the $s$-$d$ exchange
interaction $J$ is not large enough ($0<J<4/\pi^{2}$), the triplet
ground state changes from bound state to free state while varying
the magnetic flux $\phi$. However, when $J$ is large enough,
($J>4/\pi^{2}$) the ground state electron is always trapped by the
magnetic impurity at any magnetic flux $\phi$.

\begin{figure}[ptbh]
\includegraphics[width=\columnwidth]{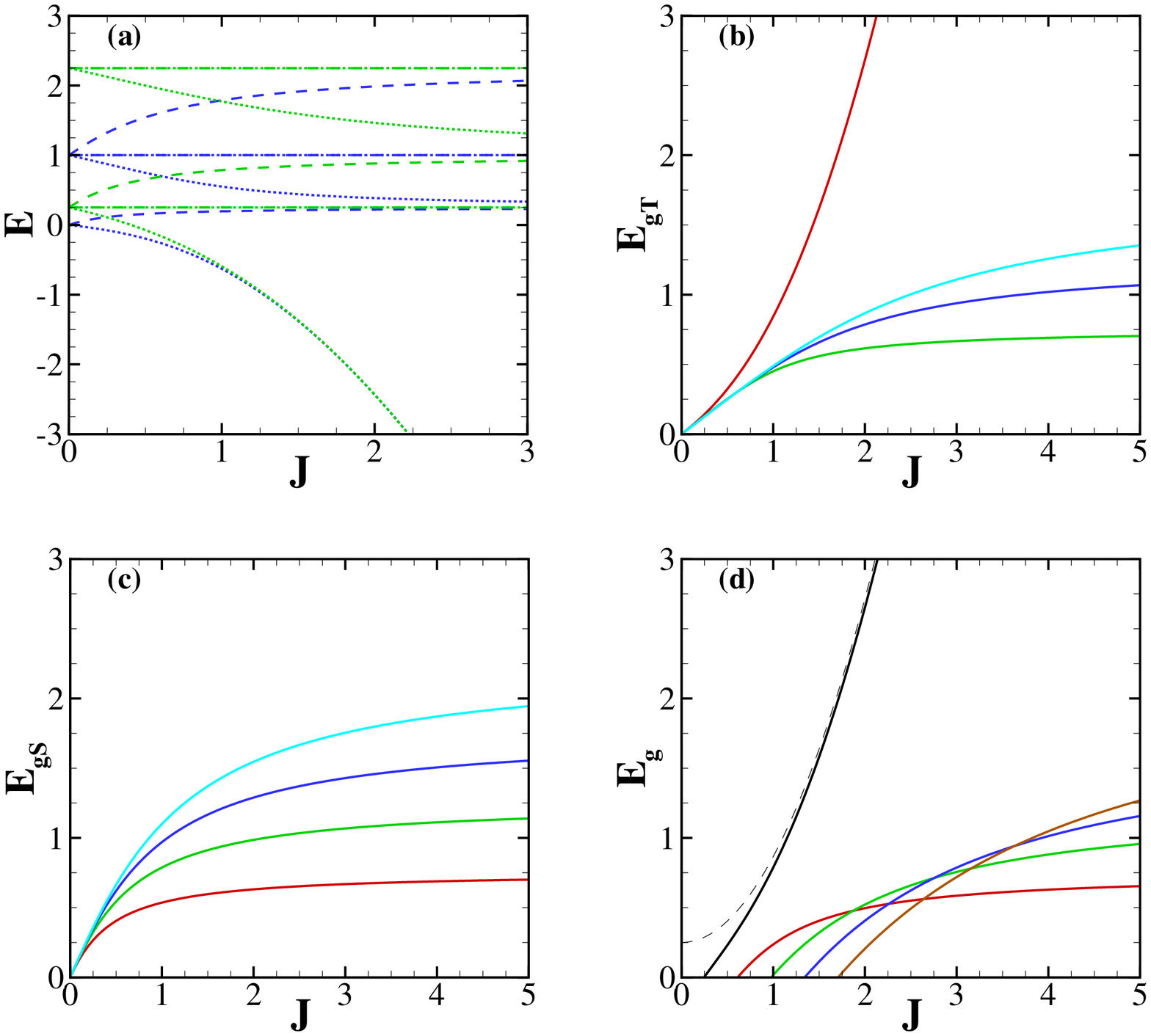}\caption{(Color
online) (a) The lowest triplet (dotted) and singlet (dashed) energy
levels as functions of the $s$-$d$ exchange interaction strength
$J$. The blue (green) lines are for the integer (half-integer)
magnetic flux $\phi$; (b) The lowest four triplet gaps as functions
of $J$, where the red (green, blue, cyan) line denotes the first
(second, third, fourth) lowest singlet gap; (c) The same as (b) but
for the singlet gaps; (d) The true energy gaps as functions of the
$s$-$d$ exchange interaction strength $J$. The dark (red, green,
blue, brown) line denotes the first (second, third, fourth, fifth)
lowest energy gap. The dashed line depicts the asymptotic behavior
of the lowest energy gap as $J$ increases. }
\label{fig:gap_appearance}
\end{figure}

Fig.~\ref{fig:gap_appearance}(a) depicts the lowest triplet (dotted)
and singlet (dashed) energy levels as functions of the strength of
the $s$-$d$ exchange interaction $J$ at an integer magnetic flux
(blue) or half-integer magnetic flux (green). The dash-dotted
horizontal lines correspond to those special fourfold degenerate
points in Fig.~\ref{fig:eigen_pd_nosoi}(a) whose energies do not
change with $J$. The energy difference of two triplet (or singlet)
states at different magnetic fluxes but of the same order approach
zero as $J \rightarrow \infty$, e.g.,
\begin{gather*}
E_{T1}(\phi=0)\rightarrow E_{T1}(\phi=1/2) \rightarrow
-\pi^{2}J^{2}/16\\
E_{S1}(\phi=0) \rightarrow E_{S1}(\phi=1/2)=1/4\\
E_{T2}(\phi=0) \rightarrow E_{T2}(\phi=1/2)=1/4\\
E_{S2}(\phi=1/2) \rightarrow E_{S2}(\phi=0)=1\\
E_{T3}(\phi=1/2) \rightarrow E_{T3}(\phi=0)=1
\end{gather*}
as $J \rightarrow \infty$. These results can be obtained from
approximate solutions to the three transcendental equations£¬i.e.,
Eqs.~(\ref{eqn:transeqns}), (\ref{eqn:transeqnt1}) and
(\ref{eqn:transeqnt2}) in the Appendix, for large $J$ at integer and
half-integer magnetic flux. The lowest four energy gaps between
triplet (singlet) states are shown in
Fig.~\ref{fig:gap_appearance}(b) [Fig.~\ref{fig:gap_appearance}(c)].
But they are only pseudo gaps. The true gaps are shown in
Fig.~\ref{fig:gap_appearance}(d), the energy gaps appear
successively and increase. The lowest energy gap
$E_{g1}=E_{S1}(\phi=0)-E_{T1}(\phi=1/2)$ approaches
$1/4+\pi^{2}J^{2}/16$ as $J \rightarrow \infty$ [see the dashed line
in Fig.~\ref{fig:gap_appearance}(d)], and the second lowest energy
gap $E_{g2}=E_{S2}(\phi=1/2)-E_{T2}(\phi=0)$ approaches $1-1/4=3/4$
as $J \rightarrow \infty$. The third, fourth, and fifth lowest
energy gaps approach $5/4$, $7/4$, and $9/4$ as $J\rightarrow
\infty$ , respectively.

\begin{figure}[ptbh]
\includegraphics[width=\columnwidth]{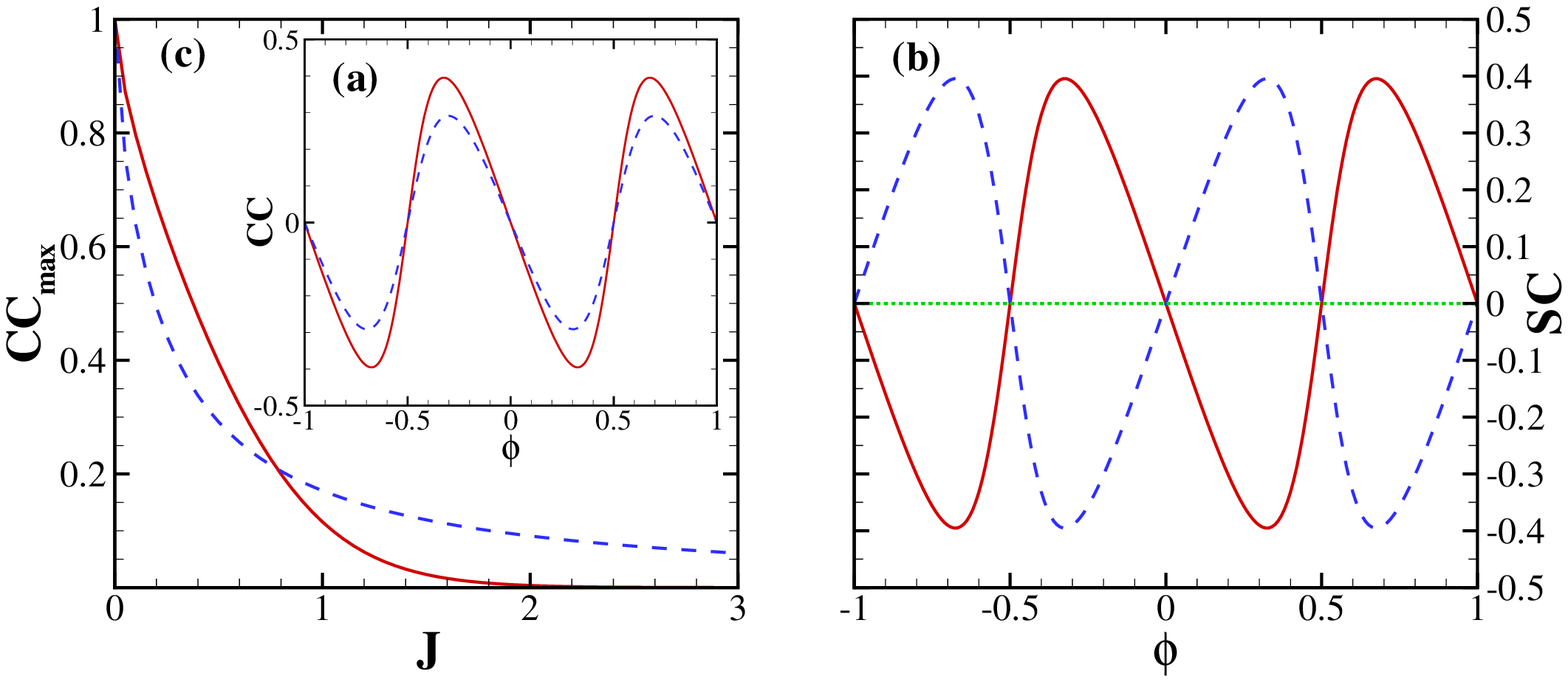}\caption{(Color online)
(a) The persistent CC from the lowest triplet (singlet) energy level
at different magnetic flux $\phi$ is denoted by the red solid (blue
dashed) line, $J=0.5$; (b) The persistent SC from the lowest $\vert
1,-1 \rangle$ ($\vert 1,1 \rangle$) state is denoted by the red
solid (blue dashed) line, and the green dotted lines are the
persistent SC from the lowest $\vert 1,0 \rangle$ state and that
from the lowest $\vert 0,0 \rangle$ state, $J=0.5$; (c) The maximal
value of the persistent CC from the lowest triplet (singlet) energy
level with different strengths of the $s$-$d$ exchange $J$ is
denoted by red solid (blue dashed) line. }
\label{fig:persistent_currents}
\end{figure}

Fig.~\ref{fig:persistent_currents}(a) shows the persistent CC's from
the lowest triplet and singlet energy levels at different magnetic
flux $\phi$. The persistent CC from both the triplet and singlet
states are smoothed and suppressed by the magnetic impurity. The
persistent SC's from the lowest $\vert 1,-1 \rangle$, $\vert 1,0
\rangle$, $\vert 1,1 \rangle$, and $\vert 0,0 \rangle$ are depicted
in Fig.~\ref{fig:persistent_currents}(b). The SC contributions of
$\vert 1,0 \rangle$ and $\vert 0,0 \rangle$ are always zero and the
SC contributions of $\vert 1,-1 \rangle$ and $\vert 1,1 \rangle$ are
always opposite, thus canceling each other. We note that the
persistent SC from the lowest $\vert 1,-1 \rangle$ is proportional
to the persistent CC from the same state. The oscillation
amplitudes, i.e., the maximal values, of the persistent CC from the
lowest triplet and singlet energy levels are shown in
Fig.~\ref{fig:persistent_currents}(c) with different strengths of
the $s$-$d$ exchange interaction $J$. The persistent CC's from both
the lowest triplet and singlet energy levels decline as $J$
increases. We recall that the magnetic impurity acts as a
$\delta$-well ($\delta$-barrier) for the triplet (singlet) states.
Both the $\delta$-well and $\delta$-barrier hinder electron
propagation along the ring and suppress the persistent CC (and SC).
The persistent CC from the lowest triplet energy level declines more
rapidly than its singlet counterpart because the electron is more
localized (around the $\delta$-well) for the triplet states than for
the singlet states. Nevertheless, for small $J$ the persistent CC
from the lowest singlet energy level can be smaller than that from
the lowest triplet energy level [see
Fig.~\ref{fig:persistent_currents}(a)] because the strength of the
$\delta$-barrier for the singlet states is triple the strength of
the $\delta$-well for the triplet states.

\begin{figure}[ptbh]
\includegraphics[width=\columnwidth]{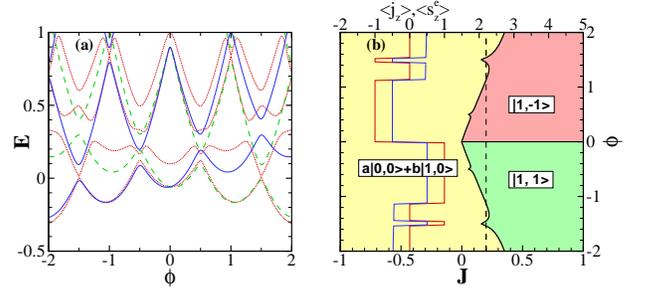}\caption{(Color online)
(a) Energy spectrum of a GaAs ring with an embedded spin-1/2
magnetic impurity ($J=0.2$), the Zeeman terms of both the electron
and the magnetic impurity are included, $g_{e}=-0.02948$ and
$g_{m}=0.134$, the red dotted lines denote those levels with zero
$\langle \hat{j}_{z}\rangle$, and the green dashed (blue solid)
lines denote the $\vert1,-1\rangle$ ($\vert1,1\rangle$) levels; (b)
The three-phase transition of the ground state, and the variations
of the $\langle \hat{j}_{z}\rangle$ (the red line) and $\langle
\hat{s}_{z}\rangle$ (the blue line) along the dashed line ($J=0.2$),
$\bar{g}_{1}=-0.02948$ and $\bar{g}_{2}=0.134$. }
\label{fig:eigen_phase}
\end{figure}

Now we include the intrinsic Zeeman terms of both the electron and
the magnetic impurity, e.g., GaAs ring. The energy spectrum with an
embedded spin-1/2 magnetic impurity is depicted in
Fig.~\ref{fig:eigen_phase}(a). The dimensionless $g$ factors
$g_{e}=g_{e}^{\ast}\*m^{\ast}=-0.02948$ and
$g_{m}=g_{m}^{\ast}\*m^{\ast}=0.134$. It is interesting to notice
that the $|0,0\rangle$ states and $|1,0\rangle$ states are coupled
together by the Zeeman terms [see Eq.~(\ref{eqn:zeemanc})]. Although
these states are mixed, the projection of the angular momentum along
the $z$-axis $\langle\hat{j}_{z}\rangle$ is still a good quantum
number, i.e., $\langle \hat{j}_{z}\rangle=0$ [see the red dotted
lines in Fig.~\ref{fig:eigen_phase}(a)]. The states $|1,-1\rangle$
and $|1,1\rangle$ are decoupled, and the Zeeman terms only alter
their energies [see the green dashed lines and the blue solid lines
in Fig. \ref{fig:eigen_phase}(a)], while the total spin
$\hat{\bm{j}}$ and its $z$-component $\hat{j}_{z}$ are still good
quantum numbers. Fig.~\ref{fig:eigen_phase}(b) shows the phase
diagram for the ground state of the ring at different $J$ and
$\phi$. From this figure one can see that the ground state in the
ring can transit among those three kinds of states due to the
interplay between the Zeeman terms and $s$-$d$ exchange interaction
as the magnetic flux $\phi$ increases, and $\langle
\hat{j}_{z}\rangle$ and $\langle \hat{s}_{z}\rangle$ undergo sudden
changes across boundaries in the phase diagram [see the red and blue
lines in Fig.~\ref{fig:eigen_phase}(b)].

\subsection{The effects of the RSOI and DSOI}

In this subsection, we focus on the competition between the $s$-$d$
exchange interaction and SOIs. From Eq.~(\ref{eqn:hami}), the
interplay between the RSOI and DSOI induces a $\sin2\varphi$
periodic potential and breaks the cylindrical symmetry of the
mesoscopic ring. The spin states, energy spectrum and persistent
currents depend sensitively on the position of the magnetic
impurity.

\begin{figure}[ptbh]
\includegraphics[width=\columnwidth]{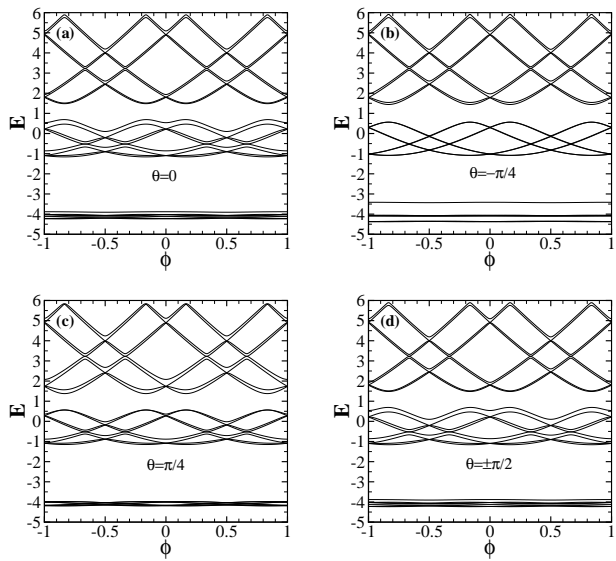}\caption{
Energy spectrum of 1D mesoscopic ring with an embedded spin-1/2
magnetic impurity and two types of SOI's, $J=0.2$, $\alpha=3$ and
$\beta=2$, the position of the impurity is $\theta=0$ in panel (a),
and $-\pi/4$, $\pi/4$, $\pm \pi/2$ in panel (b), (c), (d)
respectively. }\label{fig:eigen_soi}
\end{figure}

Fig.~\ref{fig:eigen_soi} depicts the influence of the position of
the magnetic impurity on the eigenenergy spectrum of the mesoscopic
ring in the presence of both the RSOI and DSOI. Strictly speaking,
the degeneracy of the triplet states is lifted by the RSOI and DSOI.
The position of the magnetic impurity more significantly influences
the lower energy levels than the higher levels since the
wavefunctions of the lower states are more localized than that of
the higher states. In such an anisotropic ring, $\theta=0$ and
$\theta=\pm \pi/2$ are equivalent positions which can be connected
to each other by mirror reflections with respect to the $\varphi=\pm
\pi/4$ planes, and therefore two energy spectra in panels (a) and
(d) in Fig.~\ref{fig:eigen_soi} are exactly the same. The energy
splittings due to the $s$-$d$ exchange interaction in panel (b)
[(c)] are largest (smallest) because the probability density of the
lowest bound states exhibits maxima (minima) when the magnetic
impurity is located at the bottom (peak) of the $\sin2\varphi$
periodic potential induced by the interplay between the RSOI and
DSOI. We also notice that the corresponding energy splittings of the
second bound state in panel (b) are zero because the magnetic
impurity is located just at the node ($\theta=-\pi/4$) of the
wavefunction of the second bound state.

\begin{figure}[ptbh]
\includegraphics[width=\columnwidth]{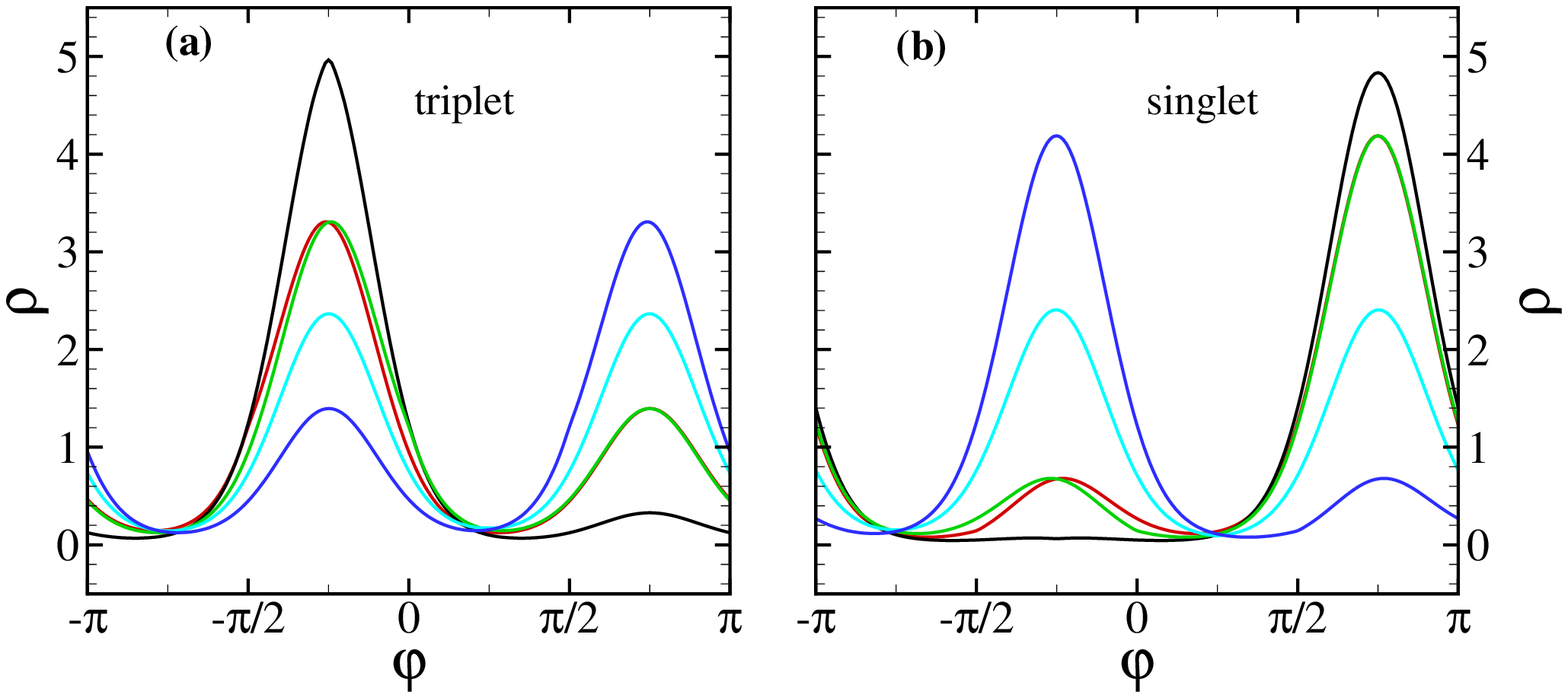}\caption{(Color online) The
probability density distributions of the lowest triplet [panel (a)]
and singlet [panel (b)] states. The red (dark, green, cyan, blue)
line is for the magnetic impurity located at $\theta=-\pi/2$ ($-\pi/
4$, $0$, $\pi/4$,$\pi /2$). Other parameters are $J=0.2$,
$\alpha=3$, $\beta=2$, and $\phi=0$. } \label{fig:pd_soi}
\end{figure}

Fig.~\ref{fig:pd_soi} shows the probability density distributions of
the lowest singlet and triplet states for different positions of the
magnetic impurity. The electron is distributed along the ring
according to the potential $\frac{\alpha \beta}{2}\sin2\varphi$,
which is induced by the interplay between the RSOI and DSOI  [see
Eq.~(\ref{eqn:hami})] in the absence of a magnetic impurity. This
potential $\frac{\alpha \beta}{2}\sin2\varphi$ exhibits two valleys
at $\varphi=-\pi/4$ and $\varphi=3\pi/4$, where the electron is most
likely to appear, and two peaks at $\varphi=\pi/4$ and
$\varphi=-3\pi/4$, corresponding to the minimum of the probability
density of electron. As shown in the previous subsection, the
magnetic impurity acts as a $\delta$-like barrier for the singlet
state electron and as a $\delta$-well for the triplet state electron
when $J>0$, and the height of the $\delta$-barrier is three time
larger than that of the $\delta$-well [see
Eq.~(\ref{eqn:singletandtriplet})]. Thus the presence of the
magnetic impurity will make the potential profile at the positions
$\varphi=-\pi/4$ and $\varphi=3\pi/4$ no longer equivalent. From
Fig.~\ref{fig:pd_soi} one can find that the competition between the
magnetic impurity and SOIs, i.e., the probability density of an
electron at the valley, is enhanced (reduced) for the triplet
(singlet) state electron when the position of the magnetic impurity
approaches the valley [see the black line in
Fig.~\ref{fig:pd_soi}(a)]. It is interesting to note that the
magnetic impurity acting as a $\delta$-like barrier for the singlet
state could also enhance the probability density of the electron at
the other valley ($\varphi=3\pi/4$) when it is at the valley
($\varphi=3\pi/4$) of the potential $\frac{\alpha \beta}{2}
\sin2\varphi$ [see the black line in Fig.~\ref{fig:pd_soi}(b)].

\begin{figure}[ptbh]
\includegraphics[width=\columnwidth]{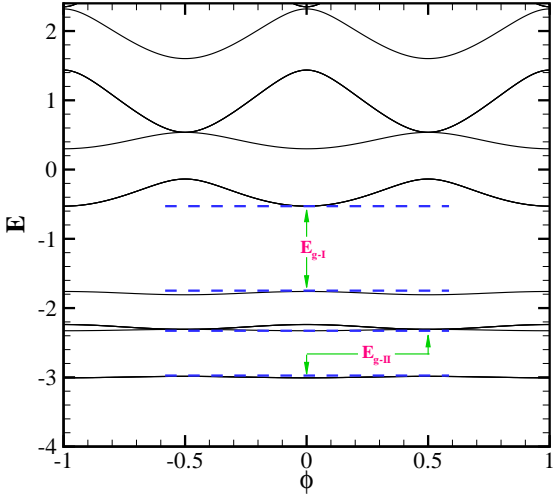}\caption{(Color online)
Two types of energy gaps appear in the energy spectrum while the
spin-orbit interactions and $s$-$d$ exchange interaction coexist in
the mesoscopic ring. $E_{\text{g-I}}$ denotes the lowest (direct)
energy gap induced by by the RSOI and DSOI, $E_{\text{g-II}}$
denotes the lowest (indirect) energy gap induced the $s$-$d$
exchange interaction. $\alpha=\beta=2$, $J=1$, $\theta=0$. }
\label{fig:gap_types}
\end{figure}

\begin{figure}[ptbh]
\includegraphics[width=\columnwidth]{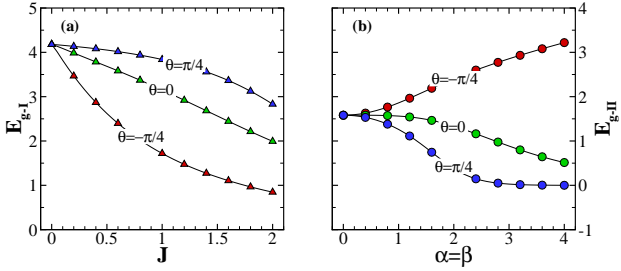}\caption{(Color online)
(a) The lowest SOI induced gap \textit{vs} the $s$-$d $ exchange
interaction strength $J$ with different positions of the magnetic
impurity, $\alpha =\beta=3$; (b) The lowest $s$-$d$ induced gap
\textit{vs} the SOI strengths ($\alpha=\beta$) with different
positions of the magnetic impurity, $J=1.5$. }
\label{fig:competition}
\end{figure}

Two types of energy gaps appear in the energy spectrum of a
mesoscopic ring, including the RSOI, the DSOI, and the $s$-$d$
exchange interaction [see Fig.~\ref{fig:gap_types}]. In our previous
work~\cite{Sheng-2006-235315}, we discussed the energy gaps caused
by the coexistence of the RSOI and DSOI ($E_{\text{g-I}}$). As shown
in Fig.~\ref{fig:gap_appearance}(d), the $s$-$d$ exchange
interaction can also open an energy gap ($E_{\text{g-II}}$) if the
strength $J$ is greater than the corresponding threshold value. We
demonstrate in Fig.~\ref{fig:competition} that the two types of
energy gaps tend to compete against each other. The lowest SOI
induced gap declines as the strength of the $s$-$d$ exchange
interaction increases [see Fig.~\ref{fig:competition}(a)]. That is
because the energy splittings between singlet and triplet states
caused by the $s$-$d$ exchange interaction tend to squeeze the gap
induced by SOI especially when the magnetic impurity approaches the
valley of potential ($\theta=-\pi/4$) since the $s$-$d$ exchange
interaction is a contact interaction that depends on the overlap
between the magnetic impurity and the electron.
Fig.~\ref{fig:competition}(b) depicts the energy gap induced by the
$s$-$d$ exchange interaction as a function of the strength of SOIs.
The increasing strengths of the RSOI and DSOI enhance the
localization of the electron. The gap increases when the magnetic
impurity is located at the valley of the potential $\frac{\alpha
\beta}{2}\sin2\varphi$ ($\theta=-\pi/4$) as the SOI strengths
($\alpha$ and $\beta$) increase, or decrease when the magnetic
impurity is at other sites. The gap width decreases most rapidly
when the magnetic impurity locates at the peak of the potential
$\frac{\alpha \beta}{2}\sin2\varphi$ ($\theta=\pi/4$).

\begin{figure}[ptbh]
\includegraphics[width=\columnwidth]{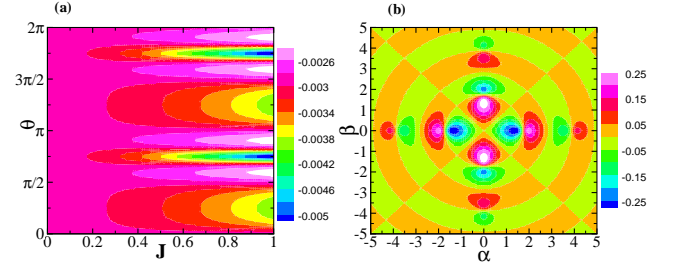}\caption{(Color online)
(a) Contour plot of the persistent SC as a function of the strength
$J$ of the $s$-$d$ exchange interaction and the impurity position
$\theta$, $\alpha=3$, $\beta=2$, and $\phi=0.5$; (b) Contour plot of
the persistent SC as a function of the RSOI strength $\alpha$ and
DSOI strength $\beta$, $J=0.5$, $\theta=0$, and $\phi=0.5$.
}\label{fig:sc_contour}
\end{figure}

The $s$-$d$ exchange interaction can also influence the persistent
SC. In Fig.~\ref{fig:sc_contour}(a), we show a contour plot of the
persistent SC as a function of the $s$-$d$ exchange interaction
strength $J$ and the position of the magnetic impurity $\theta$. We
can see that the persistent SC oscillates with the magnetic impurity
position $\theta$ when the strength $J$ is fixed. The magnitude of
the persistent SC exhibits maxima at $\theta
=\pi/4,3\pi/4,5\pi/4,7\pi/4$, where the valleys and peaks of the
potential $\frac{\alpha \beta}{2}\sin2\varphi$ are. There are also
four specific positions of magnetic impurity [the white regions in
Fig.~\ref{fig:sc_contour}(a)] where the magnitude of the persistent
SC exhibits minima. Those positions are determined by the specific
strengths of the RSOI and DSOI. When the magnetic impurity is at a
peak (valley) of the potential $\frac{\alpha \beta}{2}\sin2\varphi$,
the magnitude of the persistent SC increases (decreases) as the
$s$-$d$ exchange interaction strength $J$ increases. This provides
us a possible way to control the spin current utilizing the magnetic
impurity.

We depict the persistent SC with different RSOI strength $\alpha$
and DSOI strength $\beta$ in Fig.~\ref{fig:sc_contour}(b) at a fixed
$J$. The symmetry of the persistent SC in the $\alpha$-$\beta$
parameter space is still the same as what we reported before in the
absence of the magnetic impurity~\cite{Sheng-2006-235315}. The
eigenenergy levels become twofold degenerate when $\alpha$ and
$\beta$ are tuned to proper values in the absence of the magnetic
impurity, and the contributions from these two degenerate levels
cancel each other and consequently lead to the vanishing SC. This
twofold degeneracy will be lifted by the $s$-$d$ exchange
interaction and the the levels split into singlet and triplet
states. The contributions to the persistent SC from the singlet
states ($|0,0\rangle$) are zero while those from the triplet states
($|1,-1\rangle$, $|1,0\rangle$, $|1,1\rangle$) states cancel each
other so that the total persistent SC is still zero even in the
presence of the magnetic impurity. That is why the symmetry is
robust against the magnetic impurity. But the magnitude of the
persistent SC is suppressed by the magnetic impurity, i.e., the
$s$-$d$ exchange interaction.

\section{CONCLUSIONS}\label{sec:summary}

We have investigated theoretically the spin states and persistent
currents (CC and SC) in a 1D ring with an embedded magnetic
impurity. The $s$-$d$ exchange interaction between the electron and
the magnetic impurity splits the eigenstates into singlet states and
triplet states. The magnetic impurity acts as a $\delta$-barrier
($\delta$-well) for the singlet (triplet) states when $J>0$, opens
energy gaps in the energy spectrum, and suppresses the persistent CC
and SC. The competition between the Zeeman terms and the $s$-$d$
exchange interaction leads to a transition of the electron ground
state in the ring. The eigenenergy spectrum, probability
distribution, and persistent SC depend sensitively on the position
of the magnetic impurity. The symmetry of the persistent SC in
parameter space ($\alpha$-$\beta$) is not destroyed by the magnetic
impurity.

\begin{acknowledgments}
This work was supported by the NSFC Grant No. 60525405 and the knowledge
innovation project of CAS.
\end{acknowledgments}

\appendix*

\section{The Hamiltonian in the coupling representation}

In the decoupling representation, the basis set is the direct
product of the spin states of the electron and the magnetic impurity
$|s_{z}^{e} \rangle \bigotimes|s_{z}^{m}\rangle$. The Hamiltonian of
a one-dimensional ring with an embedded magnetic impurity can be
written as
\begin{equation}
H_{nc} =\begin{array}{r@{}cccc}
\phantom{\vert\downarrow\downarrow\rangle} &
\vert\downarrow\downarrow\rangle &
\vert\uparrow\downarrow\rangle &
\vert\downarrow\uparrow\rangle &
\vert\uparrow\uparrow\rangle
\\
\begin{array}{r}
\vert\downarrow\downarrow\rangle \\
\vert\uparrow\downarrow\rangle \\
\vert\downarrow\uparrow\rangle \\
\vert\uparrow\uparrow\rangle
\end{array}
& \left[ \begin{array}{c}
H_{-} \\
0 \\
0 \\
0
\end{array} \right.
& \begin{array}{c}
0 \\
H_{+} \\
-\pi J\delta(\varphi-\theta) \\
0
\end{array}
& \begin{array}{c}
0 \\
-\pi J\delta(\varphi-\theta) \\
H_{+} \\
0
\end{array}
& \left. \begin{array}{c}
0 \\
0 \\
0 \\
H_{-}
\end{array} \right]
\end{array},
\label{eqn:haminc}
\end{equation}
where $H_{+}=\left(  -i\frac{\partial}{\partial \varphi}+\phi
\right) ^{2}+\frac{\pi J}{2}\delta(\varphi-\theta)$ and
$H_{-}=\left(  -i\frac {\partial}{\partial \varphi}+\phi \right)
^{2}-\frac{\pi J}{2}\delta (\varphi-\theta)$. We can transform it to
the coupling representation via a unitary operator $S$ which is
related to $C$-$G$ coefficients $S_{m_{1}
m_{2}jm}^{\frac{1}{2}\frac{1}{2}}$.

\begin{gather}
H_{c}=S^{-1}H_{nc}S,\nonumber \\
S=\begin{array}{r@{}cccc}
\phantom{\vert\downarrow\downarrow\rangle}
& \vert 0,\phantom{-}0 \rangle & \vert 1,-1\rangle & \vert
1,\phantom{-}0\rangle & \vert 1,\phantom{+}1\rangle
\\
\begin{array}{r}
\vert\downarrow\downarrow\rangle \\
\vert\uparrow\downarrow\rangle \\
\vert\downarrow\uparrow\rangle \\
\vert\uparrow\uparrow\rangle
\end{array}
& \left[ \begin{array}{c}
0 \\
1/\sqrt{2} \\
-1/\sqrt{2} \\
0
\end{array} \right.
& \begin{array}{c}
1 \\
0 \\
0 \\
0
\end{array}
& \begin{array}{c}
0 \\
1/\sqrt{2} \\
1/\sqrt{2} \\
0
\end{array}
& \left. \begin{array}{c}
0 \\
0 \\
0 \\
1
\end{array} \right]
\end{array},\\
S^{-1}=S^{\prime}.\nonumber
\end{gather}

Thus the Hamiltonian in the coupling representation is
\begin{equation}
H_{c} =\begin{bmatrix} H_{S} & 0\\ 0 & H_{T}I_{3}
\end{bmatrix},
\end{equation}
where $H_{S}=\left( -i\frac{\partial}{\partial \varphi}+\phi \right)
^{2} +\frac{3\pi J}{2}\delta(\varphi-\theta)$ is the Hamiltonian for
the singlet states ($j=0$), $H_{T}=\left( -i\frac{\partial}{\partial
\varphi}+\phi \right) ^{2}-\frac{\pi J}{2}\delta(\varphi-\theta)$ is
the Hamiltonian for the triplet states ($j=1$), and $I_{3}$
represents the $3\times3$ identity matrix.

Because of the cylindrical symmetry, the position of the magnetic impurity can
be assumed to be $\theta=0$ without loss of generality.

For the singlet state, the magnetic impurity acts as a
$\delta$-potential barrier whose strength is $\frac{3\pi J}{2}$
($J>0$). All related eigenstates should be free states in the ring
($E=\kappa^{2}>0, \kappa>0$), and can be determined by
\begin{equation}
\left \{
\begin{array}[c]{l}
\left( -i\frac{\partial}{\partial \varphi}+\phi \right)^{2}X=\kappa^{2}X,0<\varphi<2\pi \\
X(0)=X(2\pi)\\
X^{\prime}(0)-X^{\prime}(2\pi)=\frac{3\pi J}{2}X(0).
\end{array}
\right. \label{eqn:solvesinglet}
\end{equation}
The general solution
$X(\varphi)=C_{1}e^{i(\kappa-\phi)\varphi}+C_{2}
e^{-i(\kappa+\phi)\varphi}$, we obtain the following transcendental
equation:
\begin{equation}
4\kappa \lbrack \cos2\pi \phi-\cos2\pi \kappa]=3\pi
J\sin2\pi\kappa.\label{eqn:transeqns}
\end{equation}
The singlet eigenenergies $E=\kappa^{2}$ can be obtained from the zeros of
Eq.~(\ref{eqn:transeqns}). We further consider the following special cases:

(a) when $\phi$ is an integer;\newline The $\kappa$ values are determined by
\begin{equation}
\sin \pi \kappa=0\  \text{or}\  \cot \pi \kappa=\frac{4\kappa}{3\pi
J} .\label{eqn:sinteger}
\end{equation}

(b) when $\phi$ is a half-integer.\newline The $\kappa$ values are determined
by
\begin{equation}
\cos \pi \kappa=0\  \text{or}\  \tan \pi \kappa=-\frac{4\kappa}{3\pi
J} .\label{eqn:shalfinterger}
\end{equation}

For the triplet state electrons, the magnetic impurity acts as a
$\delta $-potential well whose strength is $\frac{\pi J}{2}$
($J>0$). The lowest triplet state could be a bound state
($E=-\kappa^{2}<0, \kappa>0$), and the higher states can still be
free states extended over the whole ring ($E=\kappa^{2}>0,
\kappa>0$). The free triplet eigenstates can be determined by
\begin{equation}
\left \{
\begin{array}[c]{l}
\left(-i\frac{\partial}{\partial \varphi}+\phi \right)^{2}X=\kappa^{2}X,0<\varphi<2\pi \\
X(0)=X(2\pi)\\
X^{\prime}(0)-X^{\prime}(2\pi)=-\frac{\pi J}{2}X(0).
\end{array}
\right. \label{eqn:solvetriplet1}
\end{equation}
The corresponding transcendental equation can be derived similarly,
\begin{equation}
4\kappa \lbrack \cos2\pi \phi-\cos2\pi \kappa]=-\pi J\sin2\pi \kappa
.\label{eqn:transeqnt1}
\end{equation}

The eigenenergy spectrum at the specific magnetic fluxes are:

(a) when $\phi$ is an integer;\newline The $\kappa$ values are determined by
\begin{equation}
\sin \pi \kappa=0\  \text{or}\  \cot \pi \kappa=-\frac{4\kappa}{\pi
J} .\label{eqn:trinteger1}
\end{equation}

(b) when $\phi$ is a half-integer.\newline The $\kappa$ values are determined
by
\begin{equation}
\cos \pi \kappa=0\  \text{or}\  \tan \pi \kappa=\frac{4\kappa}{\pi
J} .\label{eqn:thalfinterger1}
\end{equation}

The triplet bound state can be obtained by substituting $\kappa$ with
$i\kappa$ in Eqs.~(\ref{eqn:solvetriplet1}), (\ref{eqn:transeqnt1}),
(\ref{eqn:trinteger1}), and (\ref{eqn:thalfinterger1}). The corresponding
transcendental equation is
\begin{equation}
4\kappa \lbrack \cos2\pi \phi-\cosh2\pi \kappa]=-\pi J\sinh2\pi
\kappa .\label{eqn:transeqnt2}
\end{equation}

The eigenenergy spectrum at the specific magnetic fluxes are:

(a) when $\phi$ is an integer;\newline The $\kappa$ values are determined by
\begin{equation}
\coth \pi \kappa=\frac{4\kappa}{\pi J}.\label{eqn:trinteger2}
\end{equation}

(b) when $\phi$ is a half-integer.\newline The $\kappa$ values are determined
by
\begin{equation}
\tanh \pi \kappa=\frac{4\kappa}{\pi J}.\label{eqn:thalfinterger2}
\end{equation}

The Zeeman terms are diagonal matrix elements in decoupling
representation,
\begin{equation}
H_{nc}^{Z} =
\begin{bmatrix}
-(g_{e}+g_{m})\phi & 0 & 0 & 0\\
0 &(g_{e}-g_{m})\phi & 0 & 0\\
0 & 0 & (g_{m}-g_{e})\phi & 0\\
0 & 0 & 0 & (g_{e}+g_{m})\phi
\end{bmatrix}.\label{eqn:zeemannc}
\end{equation}
But in coupling representation it becomes
\begin{equation}
\begin{split}
H_{c}^{Z} &  =S^{-1}H_{nc}^{Z}S\\
& =
\begin{bmatrix}
0 & 0 & (g_{e}-g_{m})\phi & 0\\
0 &-(g_{e}+g_{m})\phi & 0 & 0\\
(g_{e}-g_{m})\phi & 0 & 0 & 0\\
0 &0 & 0 & (g_{e}+g_{m})\phi
\end{bmatrix}.
\end{split}
\label{eqn:zeemanc}
\end{equation}
Generally speaking, $g_{e}$ is not equal to $g_{m}$, and therefore
the singlet $|0,0\rangle$ and triplet $|1,0\rangle$ states are
coupled together by the Zeeman terms (see $(g_{e}-g_{m})\phi$ in
Eq.~(\ref{eqn:zeemanc})).

\bibliography{myref}

\end{document}